\documentstyle[amssymb,aps,prl,epsfig]{revtex}

\begin{document}
\draft

\twocolumn[\hsize\textwidth\columnwidth\hsize\csname
@twocolumnfalse\endcsname

\title{Coulomb Blockade Resonances in Quantum Wires}
\author{T. Rejec$^a$ and A. Ram\v sak$^{a,b}$}
\address{$^a$J. Stefan Institute,  SI-1000 Ljubljana, Slovenia \\
$^b$Faculty of Mathematics and Physics, University of Ljubljana,
SI-1000 Ljubljana, Slovenia
}
\author{J. H. Jefferson}
\address{DERA, Electronic Sector, St. Andrews Road, Great Malvern,
Worcestershire WR14 3PS, England
}
\date{\today}
\maketitle

\begin{abstract}
\widetext
\smallskip
The conductance through a quantum wire of cylindrical cross section 
and a weak bulge is
solved exactly for two electrons within the Landauer-B\" uttiker
formalism. 
We show
that
this 'open' quantum dot exhibits spin-dependent  Coulomb blockade resonances
resulting in
two anomalous structure on the rising edge to the first conductance
plateau, one
near $0.25(2e^2/h)$, related to a singlet resonance,
and one near  $0.7(2e^2/h)$, related to a triplet resonance. 
These resonances
are generic
and robust,  occurring for other types of quantum wire and surviving to
temperatures of a few degrees.
\end{abstract}
]

\narrowtext


Recent technological advances have enabled semiconductor quantum
wires to be fabricated with effective wire widths down to a few
nanometers; for example, by heteroepitaxial growth on `v'-groove
surfaces \cite{walther92}, epitaxial growth on ridges
\cite{ramvall97}, cleaved edge over-growth \cite{yacoby96}, etched
wires with gating \cite {kristensen98}, and gated two-dimensional
electron gas (2DEG) structures \cite{wees88,wharam88}. More
recently, there has been considerable interest in carbon nanotubes
for which the quantum wire cross section can approach atomic
dimensions. Such structures have potential for opto-electronic
applications, such as light-emitting diodes, low-threshold lasers
and single-electron devices.

Many groups have now observed conductance steps in all of these
various types of quantum wire, following the pioneering work in
Refs. \cite{wees88,wharam88}. Whilst these
experiments are broadly consistent with a simple non-interacting
picture, there are certain anomalies, some of which are believed
to be related to electron-electron interactions and appear to be
spin-dependent. In particular, a structure is seen in the rising
edge of the conductance curve, starting at around $0.7(2e^{2}/h)$
and merging with the first conductance plateau with increasing
energy \cite{thomas96}. This structure, already visible in the
early experiments \cite{wees88}, can survive to temperatures of a
few degrees and also persists under increasing source-drain bias,
even when the conductance plateau has disappeared. Under
increasing in-plane magnetic field, the structure moves down,
eventually merging with the $e^{2}/h$ conductance plateau at very
high fields. Theoretical work has attempted to explain these
observation in various ways, including conductance suppression in
a Luttinger liquid with repulsive interaction and disorder
\cite{maslov95}, local spin-polarized density-functional theory
\cite{wang98} and spin-polarized subbands \cite{fasol94}. In some
of the more recent experiments, an anomaly is seen at lower energy
with conductance around $0.2(2e^{2}/h)$
\cite{kaufman99,ramvall97}. This can also survive to a few
degrees, though is less robust than the 0.7 anomaly and is more
readily suppressed by a magnetic field \cite{ramvall97}.

In this letter, we suggest that these anomalies are related to
weakly bound states and resonant bound states within the wire.
These would arise, for example, if there is a small fluctuation in
thickness of the wire in some region giving rise to a weak bulge.
If this bulge is very weak then only a single electron will be
bound. We may thus regard this system as an `open' quantum dot in
which the bound electron inhibits the transport of conduction
electrons via the Coulomb interaction. Near the conduction
threshold, there will be a Coulomb blockade and we show below that
this also gives rise to a resonance, analogous to that which
occurs in the single-electron transistor \cite{meirav91}. This is
a generic effect arising from an electron bound in some region of
the wire and such binding may arise from a number of sources,
which we do not consider explicitly. For example, in addition to a
weak thickness fluctuation, a smooth variation in confining
potential due to remote gates, contacts and depletion regions
could contribute to electron confinement along the wire or gated
2DEG. In this letter we consider only very weak confinement near
the conductance threshold for which a single electron is bound.

Consider a quantum wire of circular symmetry about the $z$-axis with
constant potential, $V(r,z)=0$ within a boundary $r_{0}(z)$ from the
symmetry axis and confining potential $V_{0}>0$ elsewhere. To be definite,
we choose parameters appropriate to GaAs for the wire and 
Al$_{x}$Ga$_{1-x}$ 
As for the barrier with $x$ such that $V_{0}=0.4$ eV, which is close to the
crossover to indirect gap. Band non-parabolicity is neglected and we use the
GaAs effective mass, $m^{\ast }=0.067m_{0}$,
neglecting its variation across
the boundary. The wire width is taken as $r_{0}(z)=\frac{1}{2}a_{0}(
1+\xi \cos ^{2}\pi z/a_{1}) $ for $|z|\leq \frac{1}{2}a_{1}$ and
$r_{0}(z)\equiv \frac{1}{2}a_{0}$ otherwise, i.e., a wire of
width $a_{0}$ with a single bulge of length $a_{1}$ and 
width $(1+\xi )a_{0}$,
as shown in Fig.~1 (inset).

\begin{figure}[htb]
\epsfig{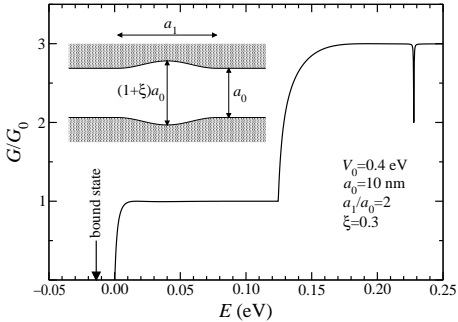} \caption{Conductance
$G/G_0$ for one (non-interacting) electron. The energy $E$ is
measured from the bottom of the lowest transverse channel. Inset:
geometry of 'open quantum dot' is determined with
$r_{0}(z)=\frac{1}{2}a_{0}( 1+\protect\xi \cos^{2}\protect\pi
z/a_{1})$.}
\end{figure}

When the wire is connected to metallic source-drain contacts,
electrons will flow into the wire region as the Fermi energy is
raised from below the conduction band edge via a gate (not
considered explicitly). At least one electron will then become
bound in the bulge region of the wire. The number of bound
electrons depends on both the Fermi energy and the relative size
of the bulge (i.e. parameters $a_1$ and $\xi $). We first consider
the non-interacting electron problem for unbound (scattering)
states. As shown in \cite{nakazato91,ramsak98} for a
two-dimensional wire, this problem may be reduced to a quasi
one-dimensional problem by expanding in transverse modes. The
Schr\"{o}dinger equation then reduces to an $N$-component
differential equation which is solved for the scattering states.
The number of channels, $N$, is chosen to be sufficiently large to
give convergence and depends upon the Fermi energy in the leads
and the dot parameter $\xi $. For $\xi $ large many channels are
needed since the rapid change in wire thickness gives rise to
large inter-channel mixing. From the solution of the scattering
problem, the conductance is calculated from the usual
Landauer-B\"{u}ttiker formula \cite{landauer57}, $G=G_{0}\;{\cal
T}(E)$, where $G_{0}=2e^{2}/h$, $E$ is the Fermi energy and ${\cal
T}(E)={\rm Tr}[{\bf t}^{\dagger }{\bf t}]$ is the total
transmission probability. This is shown in Fig.~1 for a wire with
dot parameter $\xi =0.3$. For such a small $\xi $ the conductance
is very similar to that of a perfect straight wire, as we would
expect, with conductance steps at $G_{0},3G_{0},5G_{0}$, etc. The
main difference is the very sharp Fano anti-resonance seen in the
second conductance step, a consequence of inter-channel mixing,
which is washed out at finite temperature. Apart from this
resonance, all other features, including the position of a bound
state below the conduction edge, may be accurately described by
neglecting coupling between channels. With such weak confinement,
there is only one bound state.

We now consider the interacting electron problem with wire thickness
variation in a range which ensures that only one electron occupies a bound
state and that restriction to a single channel near the conduction edge is
an excellent approximation. Within the single-channel approximation the
`few'-electron (two-electron here) problem may be reduced to an
extended Hubbard model model by discretizing the single-channel
Schr\"{o}dinger equation on a finite-difference mesh in the $z$-direction,
\begin{eqnarray}
H=- &&t\sum_{i\sigma }\left( c_{i+1,\sigma }^{\dagger }c_{i\sigma
}+c_{i\sigma }^{\dagger }c_{i+1\sigma }\right) +\sum_{i}\epsilon _{i}n_{i}+
\\ \label{hubbard}
&&+\sum_{i}U_{ii}n_{i\uparrow }n_{i\downarrow }+\frac{1}{2}\sum_{i\neq
j}U_{ij}n_{i}n_{j},  \nonumber
\end{eqnarray}
where $c_{i\sigma }^{\dagger }$ creates an electron with spin $\sigma $ at
the $z=z_{i}$ in the lowest transverse channel; $n_{i}=\sum_{\sigma
}n_{i\sigma }$ with $n_{i\sigma }=c_{i\sigma }^{\dagger }c_{i\sigma }$; $%
t=\hbar ^{2}/(2m^{\ast }\Delta ^{2})$, where $\Delta =z_{i+1}-z_{i};$ $%
\epsilon _{i}=\hbar ^{2}/(m^{\ast }\Delta ^{2})+\epsilon \left( z_{i}\right)
$, where $\epsilon \left( z_{i}\right) $ is the energy of the lowest
transverse channel at $z_{i}$. $U_{ij}$ is an
effective screened Coulomb interaction. An explicit expression
was obtained by starting with a full 3D Coulomb interaction,
and integrating over
the lowest transverse modes at $z=z_{i}$ and $z=z_{j}$. Screening
is then added phenomenologically giving the interaction,
$U_{ij}={e^{2}}/({4\pi \varepsilon \varepsilon _{0}d_{ij} })
\exp (-{\left| z_{i}-z_{j}\right| }/{\rho
})$, with $d_{ij} \to |z_i-z_j|$ for large $d_{ij}$. The dielectric
constant is taken as $\varepsilon=12.5$, appropriate for GaAs.

We emphasize that
although the Hamiltonian Eq.~(1) was derived for a certain type
of quantum wire, it is actually applicable to a much wider class of wires
and `open dot' systems, physically different cases merely modifying
the effective one-electron energies $\epsilon _{i}$ , the length
parameter $d_{ij}$ and the screening length $\rho$. The only restriction is
that the deviation from a perfectly straight wire be sufficiently small to
ensure validity of the lowest channel approximation.\

We now consider the interacting two-electron
problem in which one
electron is bound in the quantum dot region, with energy $E_{0}$.
The problem considered here is analogous to
treating the collision of an electron with a hydrogen atom as, e.g.,
in Ref.~\cite{oppenheimer28}.
It should be noted that the existence of a single-electron bound
state is guaranteed in one-dimension, at least for symmetric
wells in the $z$-direction, and in this sense is a universal
feature. With the chosen parameter range, a second electron cannot
be bound due to Coulomb repulsion. We solve the two-electron
scattering problem exactly subject to the boundary condition that
the asymptotic states consist of one bound electron in the ground
state and one free electron. From these solutions we compute the
conductance using again a Landauer-B\"{u}ttiker formula which,
incorporating the results of spin-dependent scattering
\cite{oppenheimer28}, takes the following form in zero
magnetic field,
\begin{equation}
G=G_{0}\left[ \frac{1}{4}{\cal T}_{{s}}(E)+\frac{3}{4}{\cal T}_{t}(E)\right],
\label{conductance}
\end{equation}
where the subscripts $s$ and $t$ refer to singlet and triplet configurations
respectively. In Figure~2 we show plots at zero temperature of $%
{\cal T}_{{\rm s}}(E),$ ${\cal T}_{t}(E)$ and $G/G_{0}$ for a typical wire
of width $a_{0}=10$~nm, dot width $(1+\xi )a_{0}=11.1$~nm, dot length $%
a_{1}=50$~nm, and screening lengths of 25~nm, 50~nm, and infinity. Similar
results are obtained for thicker wires, up to $a_{0}\thicksim 50$~nm, beyond
which the single-channel approximation becomes less reliable as electron
correlations become increasingly important. Note that for
weak coupling, the energy scale is set by the position of the lowest
channel, $\thicksim 1/a_{0}^{2}$ and hence the conductance vs $Ea_{0}^{2}$
is roughly independent of $a_{0}$.
\begin{figure}[htb]
\epsfig{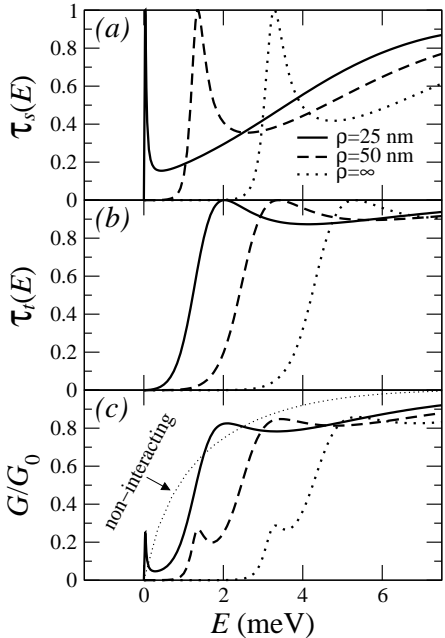} \caption{(a) Zero
temperature singlet (elastic) transmission probability ${\cal
T}_{{s}}(E)$, and (b) triplet ${\cal T}_{t}(E)$, for various
screening lengths $\rho$. The energy $E$ is defined as in Fig.~1.
(c) Total conductance, $G/G_{0}=\frac{1}{4}{\cal
T}_{{s}}(E)+\frac{3}{4}{\cal T}_{t}(E)$. Thin dotted line
represents the corresponding non-interacting result. }
\end{figure}

The main feature of these results is
that there are resonances in both singlet and triplet channels and these
give rise to structures in the rising edge to the first conductance plateau
for $G\sim \frac{1}{4}G_{0}$ (singlet) and
$G\sim \frac{3}{4}G_{0}$ (triplet).
Furthermore, as the screening is increased (screening
length reduced) we see that these resonances shift to lower energy. This
behavior has the following simple interpretation. The incident electron
feels the Coulomb potential of the electron bound in the dot region and
there is thus a gradual increase in conductance with energy, the threshold
shifting to lower energies as the screening is increased. The resonances
occur because the potential felt by the incident electron passes through
a minimum at the center of the dot, where the transverse channel energy is
lowest. Thus, the incident electron sees a double barrier which will have
some resonant energy for which there is perfect transmission. A more
detailed analysis has to account for spin and this may be understood by
gradually switching on the Coulomb interaction. For the present choice of
parameters, and also a range of parameter sets which correspond to a very
weak bulge, there are two bound states for one electron. With no interaction
both electrons may thus occupy one of 4 states (3 singlets and a triplet).
If we now switch on a small Coulomb interaction then the lowest two-electron
state will be a singlet, derived from both electrons in the lowest
one-electron state. We may regard one electron as occupying the lowest
bound-state level and the other electron of opposite spin also in this same
orbital state but energy $U$ higher, where $U$ is the intra-`atomic' Coulomb
matrix element, as in the Anderson impurity model \cite{anderson61}. As the
Coulomb interaction is increased, $U$ eventually exceeds the binding energy
and this higher level becomes a virtual bound state giving rise to a
resonance in transmission. An estimate of the energy of the virtual bound
state is given by the Anderson Coulomb matrix element with both electrons in
the one-electron orbital $\psi _{0}$, i.e., $U=\int dzdz^{\prime }|\psi
_{0}(z)|^{2}|\psi _{0}(z^{\prime })|^{2}U(z,z^{\prime })$. We have computed
this and get reasonable agreement with the exact result.

We can in addition
approximate the full scattering problem by solving the Hartree-Fock
equations without iteration in which one of the electrons again occupies $%
\psi _{0}$. The agreement is also very good and reproduces all the resonance
features. When both electrons have the same spin then they must occupy
different orbitals in the dot region when the Coulomb interaction is
switched off. With small Coulomb interaction the resulting triplet is the
lowest two-electron excited state and this develops into a resonant bound
state with the full Coulomb interaction, with energy at approximately $%
e_{1}+U_{1}-J_{1}$, where $e_{1}$\ is the energy of
the second one-electron state with $U_{1}$\ and $J_{1}$\ the respective
Coulomb and exchange integrals. The next excited (singlet) resonant bound
state is $2J_{1}$ higher in energy, which is into the first conductance
plateau region where it has little effect. We see from Fig.~2 that the
singlet resonance is somewhat sharper than the triplet and this is simply
because it is lower in energy and closer to a `real' bound state.

The resonances have a strong temperature dependence and, in particular, the
sharper singlet resonance is more readily washed out at finite temperatures.
This can be seen in Fig.~3 where we have plotted the conductance
calculated using a generalized Landauer formula
\cite{bagwell89,ramsak98} at $T=1$, 5
and 10~K for wires in which the bulge region is becoming progressively longer
and flatter, i.e., approaching a perfectly straight wire. In Fig.~3(a), with
the most pronounced deviation from a straight wire, there is only one
single-electron bound state and hence only one resonant bound state giving
rise to a peak in the conductance with $G\sim 0.3G_{0}$ at 1~K, developing
into a step at 5~K and gradually disappearing for $T>10$~K.
This behavior is expected for hard-confined wires, such as those
produced in `v'-grooves, where smooth fluctuations in thickness of
this order would be reasonable and, indeed, similar behavior has
been recently observed \cite{kaufman99}. In Fig.~3(b) the $T=0$
singlet resonance is so sharp that even at 1~K it has already
disappeared and we see only the triplet resonance which is quite
pronounced and develops into a step as the temperature is
increased, being still quite discernible at 10~K. As we progress
to a straighter wire in Fig.~3(c) we can resolve the singlet step
at 1~K but this is readily washed out as the temperature is
increased and finally for the straightest wire in Fig.~3(d), the
singlet is again unresolvable but is now swamped by the triplet
resonance, which is very close in energy. These results are
consistent with experiments on gated quantum wires and show that
only a very small deviation from a perfectly straight wire with
constant potential can give rise to the reported step-like
features near $G=0.7(2e^{2}/h).$ Furthermore, our model is
consistent with the experimental observations that these step like
anomalies will move towards a plateau at $e^{2}/h$ as the magnetic
field is increased. 
\begin{figure}[htb]
\epsfig{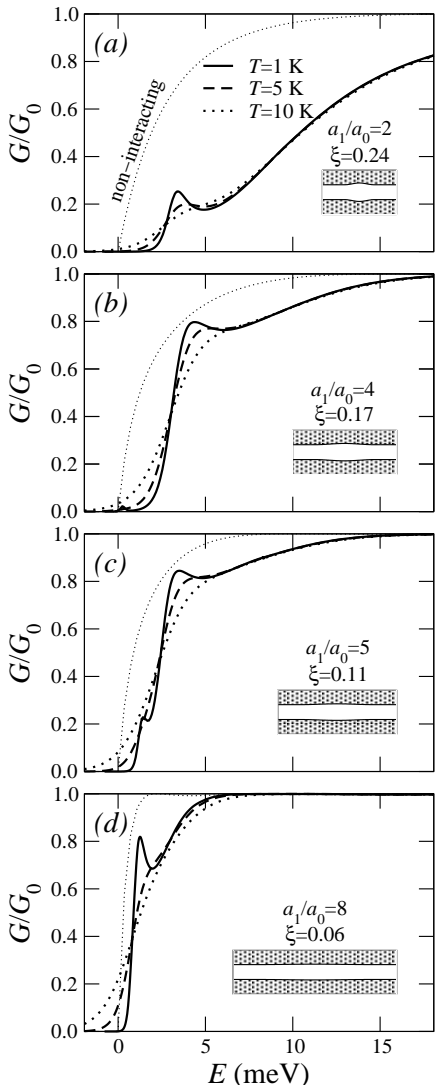}
   
\caption{Conductance $G/G_0$ for various shapes of the dot
and for various temperatures,
with screening length $\rho =50$nm.
}
\end{figure}
Further experimental observations on gated
wires \cite{thomas96,patel91} show that as the source-drain bias
is increased from zero, an anomaly appears at\ $G\thicksim
0.25(2e^{2}/h)$, coexisting with the $0.7(2e^{2}/h)$ anomaly. This
sharpens as the bias is increased and, in the example of
\cite{patel91}, for $V_{sd}\sim 6$ mV the $0.25$ anomaly is very
pronounced whilst the $0.7$ anomaly has disappeared. This is also
consistent with the above model since under bias the triplet
resonant bound-state will eventually disappear because the
confining potential in the $z$-direction will only accommodate a
single one-electron bound state, giving rise to a singlet
resonance only. Furthermore, this resonance will become broader
with increasing bias resulting in a more pronounced step, as
observed. Whilst the gross features of many of the observed
results may be interpreted in terms of this simple model of
spin-dependent scattering from a single-bound electron, the
observation of anomalies in a wide variety of samples would
mitigate against the shallow longitudinal confinement potential as
always arising from thickness fluctuations or depletion charge. A
further possibility is that a weak potential well arises
spontaneously in the Luttinger liquid due to exchange-correlation
effects in an otherwise perfect wire \cite{Khaetskii}. A treatment
of this is, however, beyond the scope of the present work.

In conclusion, we have shown that quantum wires with weak
longitudinal confinement can give rise to spin-dependent, Coulomb
blockade resonances when a single electron is bound in the
confined region, a universal effect in one-dimensional systems
with very weak longitudinal confinement. The positions of the
resulting features at $G\sim \frac{1}{4}G_{0}$ and $G\sim
\frac{3}{4}G_{0}$ are a consequence of the singlet and triplet
nature of the resonances. Further experiments in which the widths
of quantum wires and/or the confinement potentials are engineered
to control longitudinal confinement should throw further light on
the problem of spin-dependent ballistic transport.

The authors wish to acknowledge K. J. Thomas, A. V. Khaetskii, C. J.
Lambert and M. Pepper for helpful comments. This work was
supported by the EU and the MoD.

\end{document}